\documentclass[12pt]{iopart}

\pdfoutput=1\usepackage{amssymb}
\usepackage{amsthm}
\usepackage{graphicx}
\usepackage[all]{xy}

\newcommand{\bra}[1]{\langle #1 |}
\newcommand{\ket}[1]{| #1 \rangle}

\newcommand{\proj}[1]{\ket{#1}\bra{#1}}

\newcommand{\eqref}[1]{\ref{#1}}

\newcommand{\one}{{\bf 1}}

\newcommand{\prlsection}[1]{\section{#1}}

\newcommand{\chan}[1]{\mathcal #1}

\theoremstyle{plain}

\begin{document}

\title{Causal structure of the entanglement renormalization ansatz}
\author{C\'edric B\'eny}
\address{Institut f\"ur Theoretische Physik, Leibniz Universit\"at Hannover, Appelstra{\ss}e 2, 30167 Hannover, Germany}
\date{August 21, 2012}

\begin{abstract}
We show that the multiscale entanglement renormalization ansatz (MERA) can be reformulated in terms of a causality constraint on discrete quantum dynamics. This causal structure is that of de Sitter space with a flat spacelike boundary, where the volume of a spacetime region corresponds to the number of variational parameters it contains. This result clarifies the nature of the ansatz, and suggests a generalization to quantum field theory. It also constitutes an independent justification of the connection between MERA and hyperbolic geometry which was proposed as a concrete implementation of the AdS-CFT correspondence.  
\end{abstract}

\maketitle

The multiscale entanglement renormalization ansatz (MERA), introduced in \cite{vidal08}, has been successfully used to model the physics of many low-dimensional strongly correlated quantum many-body systems~\cite{vidal07,evenbly09b,giovannetti09,vidal09,cincio08,evenbly09,evenbly10,corboz09,pineda10}, thanks to its ability to simultaneously represent correlations at widely different length scales. It is defined as the set of states which can be output by a quantum circuit with fixed input and a given gate structure. The components of the gates are the variational parameters.
 An essential feature of the circuit that makes up a MERA is its particular causal structure which allows for the efficient evaluation of the expectation value of local observables~\cite{vidal08}.  

In this paper, we show that, in fact, the causal structure alone---defined as a constraint on the flow of information within each computational step---is sufficient to define the ansatz. 
By clarifying how and when the specific details of the circuit are unimportant, this approach unifies and simplifies the formulations of a MERA, and opens the way to a deeper theoretical study of the ansatz, such as its relation to other theories of renormalization~\cite{beny12}. 

More precisely, we show that the variational class of states that is a MERA can be conceived as the set of possible states that is computed by any discrete process satisfying certain specific causality constraints. This point of view naturally suggests a continuum analogy. Indeed, a set of causality constraints on a continuous dynamics is essentially a Lorentzian metric. The set of processes satisfying the constraints are then to be understood as possible quantum matter fields living in that spacetime.

Following this observation we show that MERA, expressed as a set of causality constraints, corresponds to a discretization of {\em de Sitter space}:
a solution of Einstein's equation describing an exponentially expanding universe. 
This result constitutes a connection between MERA and hyperbolic geometry which appears complementary to arguments previously proposed in connection with the AdS-CFT correspondence~\cite{swingle09,evenbly11}\footnote{Hyperbolic space is the Euclidean form of both anti de Sitter and de Sitter spacetimes.}. 

In addition, this observation naturally points to quantum field theory (QFT) on de Sitter space as a continuous generalization of MERA. We show that this proposal is compatible with the cMERA introduced in Ref.~\cite{haegeman11} as a variational ansatz for quantum field theories, but more constrained, and amenable to the tools developed in the context of quantum cosmology. 
The connection with de Sitter space is interesting for two reasons. Firstly, QFT on de Sitter background has been extensively studied given that it models the early inflationary universe where quantum effects are believed to be important. Secondly, it is arguably the next simplest spacetime in which to do QFT after Minkowski space due to its maximum number of symmetries.

\prlsection{A quick introduction on quantum channels} 
A conveniently general type of ``process'' in quantum mechanics is formalized by {\em quantum channels}, or trace-preserving and completely positive maps. They are defined as the most general maps from density matrices to density matrices which are compatible with the probability interpretation of the convex combination of states, and stable under the tensor product. They describe the evolution of open systems in the sense that a channel $\mathcal E$, acting on the (mixed) state $\rho_A$ of a system $A$, can always be written as
\[
\mathcal E(\rho) = \tr_{B} [U (\rho_A \otimes \proj 0_B) U^\dagger]
\]
where $U$ is a unitary map describing the joint evolution (interaction) of $A$ together with an auxiliary system $B$ for a fixed amount of time, the state $\ket 0_B$ is some arbitrary initial state of $B$, and $\tr_B$ is the partial trace over system $B$. Since a redefinition of $\ket 0$ can be easily canceled by a redefinition of $U$, it is often convenient to simply write $\mathcal E(\rho) = \tr_B V \rho_A V^\dagger$ where $V: \ket \psi_A \mapsto U(\psi_A \otimes \ket 0_B)$ is an {\em isometry} from system $A$ to system $AB$. More generally, an isometry is any operator $V$ satisfying $V^\dagger V = \one$ (which implies that $V V^\dagger$ is a projector), which is all we need to produce a quantum channel. 

Just as for unitary evolutions, quantum channels can be formulated in the Heisenberg picture, defined by the duality
\[
\tr(\mathcal E(\rho) X) = \tr(\rho\, \mathcal E^\dagger(X))
\]
where $\rho$ is any state and $X$ any observable. If $\mathcal E$ is defined in terms of the isometry $V$ as above, then
\[
\mathcal E^\dagger(X_A) = V^\dagger (X_A \otimes \one_B) V,
\]
in which we see that we obtain a channel by restricting observations of the state $V \rho V^\dagger$ of system $AB$ to observables of system $A$ only. This is how channels naturally appear in this work, and generally when evaluating a MERA on a local observable. 

Most properties of channels can be derived from the Stinespring dilation theorem which implies the existence of $V$ for any channel as well as the uniqueness of $V$ up to any partial isometry on the auxiliary system $B$, i.e., two isometries $V$ and $V'$ producing the same channel must satisfy $(\one_A \otimes W_B) V = V'$, where $W^\dagger_B W_B$, and hence also $W_B W_B^\dagger$, are projectors. 

\prlsection{MERA from causal order}

One step of a circuit defining a MERA takes a quantum state defined on a coarse-grained lattice and isometrically maps it into the larger Hilbert space of a finer lattice. 
These isometric steps are also required to be implemented through local gates and a fixed number of computational steps (see for instance the diagrams in Ref.~\cite{vidal08}). This implies a finite speed of information propagation in the circuit. Combined with the exponential nature of the successive coarse-graining operations, this means that the expectation value of a local observable can be evaluated in a time logarithmic in the lattice size. In order to see this, note that an expectation value can be evaluated by evolving the observable ``back in time'' in the Heisenberg picture and then computing the expectation value between the initial fiducial state and the resulting observable. When talking about locality it is enlightening to adopt the Heisenberg picture because there exists an unambiguous concept of a local observable: one which acts nontrivially only on certain lattice sites. The locality of the isometries then implies that an observable local to a region $\Sigma$ of the lattice is pulled back to an observable which is itself local with respect to a region $\Sigma'$ that is in the {\em causal past} of $\Sigma$~\cite{vidal08,evenbly09b}. This map is the dual of a local quantum channel mapping states defined on $\Sigma'$ to states on $\Sigma$. In performing this operation, the rest of the isometry can be completely ignored. Furthermore, the coarse-graining is such that the Hilbert space dimension associated with the causal past $\Sigma'$ is no larger than that of $\Sigma$, and hence the computational load can only decrease at each step, and is independent of the lattice size.

This suggests that, in defining the ansatz, we could replace the ad-hoc requirement that each isometric step have a particular gate structure, and just require it to pull back local observables (on $\Sigma$) to local observables (on $\Sigma'$). This property is precisely one of {\em causality} as it is equivalent to stating that the degrees of freedom outside $\Sigma'$ cannot influence those inside $\Sigma$ through one step of the dynamics~\cite{beckman01}. But is such a property sufficient to obtain an efficient local parameterization of the isometries? 

It was shown by Arrighi {\it et al.}~\cite{arrighi10} in the context of a unitary dynamical step $U$ that such causal constraints are sufficient and necessary for $U$ to be implementable as a circuit of local operations, or ``gates'', with some commutativity constraints between them.
In order to see that this result is non-trivial, it helps considering the fact that it fails even when the dynamical step is isometric rather than unitary. Indeed, an isometry can always acausally produce a state correlated over arbitrary distances, something which cannot be achieved with local gates and a fixed number of computational steps. Consider for instance the isometry $V$ from two distant systems $A$ and $B$ to the extended systems $A = A_1A_2$ and $B = B_1 B_2$ defined by 
\[
V( \ket{\psi}_{A_1} \otimes \ket{\phi}_{B_1}) = U_{A}\otimes U_{B} ( \ket{\psi}_{A_1} \otimes \ket{\Omega}_{A_2 B_2} \otimes \ket{\phi}_{B_1})
\]
where $\Omega$ is an entangled state and $U_{A}$ and $U_{B}$ are unitary operators acting respectively on systems $A$ and $B$.
It is enlightening to rewrite this as the circuit
\[
 \xy(0,0)*+{\xy(-4,0)*{};(-4,4)*{}**\dir{-};(-6,2)*{\scriptstyle A_1};(4,0)*{};(4,4)*{}**\dir{-};(6,2)*{\scriptstyle B_1};(-6,4)*{};(-6,16)*{}**\dir{-};(-6,16)*{};(6,16)*{}**\dir{-};(6,4)*{};(6,16)*{}**\dir{-};(-6,4)*{};(6,4)*{}**\dir{-};(0,10)*{V};(-4,16)*{};(-4,20)*{}**\dir{-};(-6,18)*{\scriptstyle A};(4,16)*{};(4,20)*{}**\dir{-};(6,18)*{\scriptstyle B};\endxy};\endxy = \xy(0,0)*+{\xy(0,1.6)*{\Omega};(-6,4)*{};(6,4)*{}**\crv{(-6,-4)&(6,-4)};(-6,4)*{};(6,4)*{}**\dir{-};(-8,5)*{};(-8,8)*{}**\dir{-};{\ar(-8,0)*{};(-8,5)*{}};(-10,4)*{\scriptstyle A_1};(-4,4)*{};(-4,8)*{}**\dir{-};(-10,8)*{};(-10,16)*{}**\dir{-};(-10,16)*{};(-2,16)*{}**\dir{-};(-2,8)*{};(-2,16)*{}**\dir{-};(-10,8)*{};(-2,8)*{}**\dir{-};(-6,12)*{{U_{A}}};(4,4)*{};(4,8)*{}**\dir{-};(8,5)*{};(8,8)*{}**\dir{-};{\ar(8,0)*{};(8,5)*{}};(10,4)*{\scriptstyle B_1};(2,8)*{};(2,16)*{}**\dir{-};(2,16)*{};(10,16)*{}**\dir{-};(10,8)*{};(10,16)*{}**\dir{-};(2,8)*{};(10,8)*{}**\dir{-};(6,12)*{{U_{B}}};(-8,16)*{};(-8,20)*{}**\dir{-};(-10,18)*{\scriptstyle A};(8,16)*{};(8,20)*{}**\dir{-};(10,18)*{\scriptstyle B};\endxy};\endxy\]
where the time flows upward. 
It is not hard to see that this setup does not allow for any communication at all from $A_1$ to $B$ nor from $B_1$ to $A$. But despite this absence of cross communication, this isometry cannot be broken down as a product $V = V_1 \otimes V_2$ where $V_1$ maps $A_1$ to $A$ and $V_2$ maps $B_1$ to $B$. 
We note that $V$ cannot be unitary in this example because the output dimension must be larger than the input dimension. 
Counter examples can become much more intricate if the transformation is implemented by a generic channel rather than an isometry~\cite{beckman01}, or in fact even by a {\em classical} channel (i.e. one mapping diagonal density matrices to diagonal density matrices)~\cite{arrighi11}. 

For a MERA, the dynamical steps cannot be unitary since the Hilbert space dimension must increase. In fact, one can show that even for an infinite lattice, the causal structure of MERA is incompatible with unitarity. However, we cannot allow for the circuit to create arbitary correlated states over long distances, even if those are irrelevant for the calculation of local expectation values, because this would lead to a number of variational parameters exponential in the lattice site.  
This means that we 
need to introduce a stronger requirement which forces even isometric maps to be locally implementable. Because it is a worthwhile generalization, we will also demand of our new constraints that they force the localization of generic quantum channels. 

In the following we will always work with respect to some causal relation between the multiple input and output systems of a map. That is, a map (in general a channel) has a set of input vertices each associated with a Hilbert space, and a possibly different set of output vertices, each also associated with a Hilbert space. The causal relation then is specified as a set of pairs of input and output vertices, i.e., a bipartite graph. We will say that the map is {\em causal} (with respect to this causal relation, or graph), if no information is transmitted between pairs which are {\em not} in the graph. Hence the graph contains those pairs which are allowed to communicate. 

As we have seen, this concept of causality is not strong enough for our purpose. Therefore we will define a new concept, that of {\em pure causality}. As explained in the introduction on quantum channels, a channel can always be written in terms of a unitary map on a larger space, i.e., with a larger input system and a larger output system. Note that if the channel is already isometric, i.e., of the form $\mathcal E(\rho) = V \rho V^\dagger$ for an isometry $V$, then we only need to enlarge the input space (so as to make it of the same dimension as the output space). 
Since the concept of causality is sufficient to enforce the locality of a unitary map, we will simply say that a channel is {\em purely causal} (with respect to a given graph) if it can be expressed in terms of a unitary map which is causal (with respect to that same graph), where the extra input or output systems are distributed among input and output vertices, and the input state on the extra inputs has no correlations. Hence the unitary is defined on the same input and output lattices as the channel, but each vertices is associated with a possibly larger Hilbert space. Since the operation of initializing and tracing out the extra local spaces are local, it is straightforward to see that a local implementation for the unitary map yields a local implementation for the corresponding channel.

For instance, consider a channel $\mathcal E$ from systems $A$ and $B$ to systems $X$ and $Y$, and the causal relation which only allows for communication from $A$ to $X$ and from $B$ to $Y$, i.e., defined by the pairs $(A,X)$ and $(B,Y)$, then $\mathcal E$ is purely causal with respect to that causality relation if there exists systems $A'$, $B'$, $X'$ and $Y'$, and a unitary operator $U$ from $AA'BB'$ to $XX'YY'$ causal with respect to the causality relation defined by the pairs $(AA',XX')$ and $(BB', YY')$, and states $\ket 0_{A'}$ and $\ket 0_{B'}$, such that 
\[
\mathcal E(\rho_{AB}) = \tr_{X'Y'}U(\rho_{AB}\otimes \proj 0_{A'} \otimes \proj 0_{B'}) U^\dagger.
\]  

We will now prove that one step of the binary MERA, defined as the {\em set} of all possible isometries which can be implemented with the gate structure of one MERA step (by picking the right parameters for the gates) is equal to the set of isometries which are purely causal with respect to a certain causality relation. Namely, that,
\begin{eqnarray}
\label{step}
\xy(0,0)*+{\xy(-12.5,-2.5)*\xycircle(0.5,0.5){}*\frm{*};(-7.5,-2.5)*\xycircle(0.5,0.5){}*\frm{*};(-2.5,-2.5)*\xycircle(0.5,0.5){}*\frm{*};(2.5,-2.5)*\xycircle(0.5,0.5){}*\frm{*};(7.5,-2.5)*\xycircle(0.5,0.5){}*\frm{*};(12.5,-2.5)*\xycircle(0.5,0.5){}*\frm{*};(-17.5,-2.5)*{};(-17.5,-2.5)*{}**\dir{-};(-16.25,-6.25)*{};(-12.5,-2.5)*{}**\dir{-};{\ar(-17.5,-7.5)*{};(-15.275,-5.275)*{}};(-10,-10)*\xycircle(0.5,0.5){}*\frm{*};(-13.75,-6.25)*{};(-17.5,-2.5)*{}**\dir{-};{\ar(-10,-10)*{};(-14.725,-5.275)*{}};(-11.25,-6.25)*{};(-12.5,-2.5)*{}**\dir{-};{\ar(-10,-10)*{};(-11.575,-5.275)*{}};(-8.75,-6.25)*{};(-7.5,-2.5)*{}**\dir{-};{\ar(-10,-10)*{};(-8.425,-5.275)*{}};(-6.25,-6.25)*{};(-2.5,-2.5)*{}**\dir{-};{\ar(-10,-10)*{};(-5.275,-5.275)*{}};(0,-10)*\xycircle(0.5,0.5){}*\frm{*};(-3.75,-6.25)*{};(-7.5,-2.5)*{}**\dir{-};{\ar(0,-10)*{};(-4.725,-5.275)*{}};(-1.25,-6.25)*{};(-2.5,-2.5)*{}**\dir{-};{\ar(0,-10)*{};(-1.575,-5.275)*{}};(1.25,-6.25)*{};(2.5,-2.5)*{}**\dir{-};{\ar(0,-10)*{};(1.575,-5.275)*{}};(3.75,-6.25)*{};(7.5,-2.5)*{}**\dir{-};{\ar(0,-10)*{};(4.725,-5.275)*{}};(10,-10)*\xycircle(0.5,0.5){}*\frm{*};(6.25,-6.25)*{};(2.5,-2.5)*{}**\dir{-};{\ar(10,-10)*{};(5.275,-5.275)*{}};(8.75,-6.25)*{};(7.5,-2.5)*{}**\dir{-};{\ar(10,-10)*{};(8.425,-5.275)*{}};(11.25,-6.25)*{};(12.5,-2.5)*{}**\dir{-};{\ar(10,-10)*{};(11.575,-5.275)*{}};(13.75,-6.25)*{};(17.5,-2.5)*{}**\dir{-};{\ar(10,-10)*{};(14.725,-5.275)*{}};(16.25,-6.25)*{};(12.5,-2.5)*{}**\dir{-};{\ar(17.5,-7.5)*{};(15.275,-5.275)*{}};(17.5,-2.5)*{};(17.5,-2.5)*{}**\dir{-};\endxy};\endxy\;\;=\;
\cdots \xy(0,0)*+{\xy(-8,3)*{};(-8,4)*{}**\dir{-};{\ar(-8,0)*{};(-8,3)*{}};(-11,4)*{};(-11,10)*{}**\dir{-};(-11,10)*{};(-5,10)*{}**\dir{-};(-5,4)*{};(-5,10)*{}**\dir{-};(-11,4)*{};(-5,4)*{}**\dir{-};(-8,7)*{{}};(0,3)*{};(0,4)*{}**\dir{-};{\ar(0,0)*{};(0,3)*{}};(-3,4)*{};(-3,10)*{}**\dir{-};(-3,10)*{};(3,10)*{}**\dir{-};(3,4)*{};(3,10)*{}**\dir{-};(-3,4)*{};(3,4)*{}**\dir{-};(0,7)*{{}};(8,3)*{};(8,4)*{}**\dir{-};{\ar(8,0)*{};(8,3)*{}};(5,4)*{};(5,10)*{}**\dir{-};(5,10)*{};(11,10)*{}**\dir{-};(11,4)*{};(11,10)*{}**\dir{-};(5,4)*{};(11,4)*{}**\dir{-};(8,7)*{{}};(2,10)*{};(2,12)*{}**\dir{-};(6,10)*{};(6,12)*{}**\dir{-};(1,12)*{};(1,18)*{}**\dir{-};(1,18)*{};(7,18)*{}**\dir{-};(7,12)*{};(7,18)*{}**\dir{-};(1,12)*{};(7,12)*{}**\dir{-};(4,15)*{{}};(-2,10)*{};(-2,12)*{}**\dir{-};(-6,10)*{};(-6,12)*{}**\dir{-};(-7,12)*{};(-7,18)*{}**\dir{-};(-7,18)*{};(-1,18)*{}**\dir{-};(-1,12)*{};(-1,18)*{}**\dir{-};(-7,12)*{};(-1,12)*{}**\dir{-};(-4,15)*{{}};(-10,16)*{};(-10,20)*{}**\dir{-};{\ar(-10,10)*{};(-10,16)*{}};(-6,18)*{};(-6,20)*{}**\dir{-};(-2,18)*{};(-2,20)*{}**\dir{-};(2,18)*{};(2,20)*{}**\dir{-};(6,18)*{};(6,20)*{}**\dir{-};(10,16)*{};(10,20)*{}**\dir{-};{\ar(10,10)*{};(10,16)*{}};\endxy};\endxy \cdots
\end{eqnarray}
where time flows upward. The left-hand side represents the set of isometries between the two one-dimensional lattices whose vertices are represented as dots (with some arbitrarily fixed Hilbert space associated with each dot) which are purely causal with respect to the causality relation specified by the graph (i.e. communications is only allowed between connected dots). The right-hand side of the equation represents the set of isometries which can be implemented by the circuit (or tensor network) obtained by replacing each box by an isometry. This is precisely one step of the binary MERA \cite{vidal08,evenbly09b}, except for the fact that in our case there is no constraint on the dimension of the Hilbert spaces associated with the intermediate wires. However, the fact that each box must be an isometry effectively limits their input of dimension to that of their output. 

Below we also show that the ternary MERA~\cite{evenbly09b} is equivalent to such a natural causality constraint.
More generally, our prescription together with the constructive localizability result introduced in Ref.~\cite{arrighi10} allows for the construction of circuits with equivalent properties on arbitrary lattices, including lattices embedded in higher dimensional spaces. 

Our approach works just as well if we allow each step to be implemented by a quantum channel rather than just an isometry, hence allowing in principle for the characterization of mixed states with long range correlations, such as critical thermal states.  
If the state to be described is classical one may furthermore constrain the local quantum channels to be stochastic maps (i.e. to map diagonal matrices to diagonal matrices).

\prlsection{Connection with de Sitter space}
Let us call an ``event'' a lattice site at a given coarse-graining step. Each event is associated with the Hilbert space dimension of the corresponding lattice site. The causal relations between successive coarse-graining generates a partial order between any two events, i.e., $A<B$ if $A$ is in the causal past of $B$. Furthermore the original causal relation can be recovered uniquely from the partial order by noting that it is defined by the causal {\em links}: pairs of related events with no events ``in between'', i.e., $(A,B)$ forms a link if $A < B$ and there is no $C$ such that $A<C<B$. 

\begin{figure}
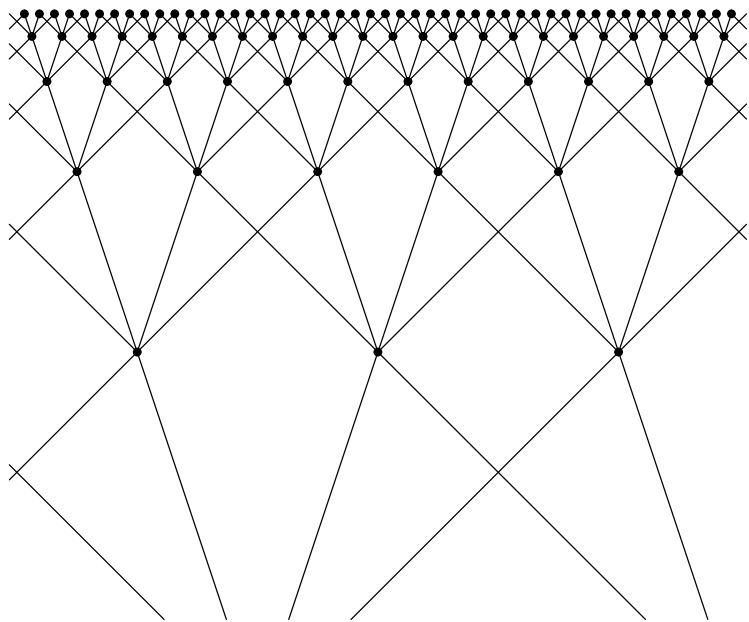

\[
\xy(0,0)*+{\xy(-47,-1)*\xycircle(0.5,0.5){}*\frm{*};(-45,-1)*\xycircle(0.5,0.5){}*\frm{*};(-43,-1)*\xycircle(0.5,0.5){}*\frm{*};(-41,-1)*\xycircle(0.5,0.5){}*\frm{*};(-39,-1)*\xycircle(0.5,0.5){}*\frm{*};(-37,-1)*\xycircle(0.5,0.5){}*\frm{*};(-35,-1)*\xycircle(0.5,0.5){}*\frm{*};(-33,-1)*\xycircle(0.5,0.5){}*\frm{*};(-31,-1)*\xycircle(0.5,0.5){}*\frm{*};(-29,-1)*\xycircle(0.5,0.5){}*\frm{*};(-27,-1)*\xycircle(0.5,0.5){}*\frm{*};(-25,-1)*\xycircle(0.5,0.5){}*\frm{*};(-23,-1)*\xycircle(0.5,0.5){}*\frm{*};(-21,-1)*\xycircle(0.5,0.5){}*\frm{*};(-19,-1)*\xycircle(0.5,0.5){}*\frm{*};(-17,-1)*\xycircle(0.5,0.5){}*\frm{*};(-15,-1)*\xycircle(0.5,0.5){}*\frm{*};(-13,-1)*\xycircle(0.5,0.5){}*\frm{*};(-11,-1)*\xycircle(0.5,0.5){}*\frm{*};(-9,-1)*\xycircle(0.5,0.5){}*\frm{*};(-7,-1)*\xycircle(0.5,0.5){}*\frm{*};(-5,-1)*\xycircle(0.5,0.5){}*\frm{*};(-3,-1)*\xycircle(0.5,0.5){}*\frm{*};(-1,-1)*\xycircle(0.5,0.5){}*\frm{*};(1,-1)*\xycircle(0.5,0.5){}*\frm{*};(3,-1)*\xycircle(0.5,0.5){}*\frm{*};(5,-1)*\xycircle(0.5,0.5){}*\frm{*};(7,-1)*\xycircle(0.5,0.5){}*\frm{*};(9,-1)*\xycircle(0.5,0.5){}*\frm{*};(11,-1)*\xycircle(0.5,0.5){}*\frm{*};(13,-1)*\xycircle(0.5,0.5){}*\frm{*};(15,-1)*\xycircle(0.5,0.5){}*\frm{*};(17,-1)*\xycircle(0.5,0.5){}*\frm{*};(19,-1)*\xycircle(0.5,0.5){}*\frm{*};(21,-1)*\xycircle(0.5,0.5){}*\frm{*};(23,-1)*\xycircle(0.5,0.5){}*\frm{*};(25,-1)*\xycircle(0.5,0.5){}*\frm{*};(27,-1)*\xycircle(0.5,0.5){}*\frm{*};(29,-1)*\xycircle(0.5,0.5){}*\frm{*};(31,-1)*\xycircle(0.5,0.5){}*\frm{*};(33,-1)*\xycircle(0.5,0.5){}*\frm{*};(35,-1)*\xycircle(0.5,0.5){}*\frm{*};(37,-1)*\xycircle(0.5,0.5){}*\frm{*};(39,-1)*\xycircle(0.5,0.5){}*\frm{*};(41,-1)*\xycircle(0.5,0.5){}*\frm{*};(43,-1)*\xycircle(0.5,0.5){}*\frm{*};(45,-1)*\xycircle(0.5,0.5){}*\frm{*};(47,-1)*\xycircle(0.5,0.5){}*\frm{*};(-49,-1)*{};(-49,-1)*{}**\dir{-};(-49,-3)*{};(-47,-1)*{}**\dir{-};(-46,-4)*\xycircle(0.5,0.5){}*\frm{*};(-46,-4)*{};(-49,-1)*{}**\dir{-};(-46,-4)*{};(-47,-1)*{}**\dir{-};(-46,-4)*{};(-45,-1)*{}**\dir{-};(-46,-4)*{};(-43,-1)*{}**\dir{-};(-42,-4)*\xycircle(0.5,0.5){}*\frm{*};(-42,-4)*{};(-45,-1)*{}**\dir{-};(-42,-4)*{};(-43,-1)*{}**\dir{-};(-42,-4)*{};(-41,-1)*{}**\dir{-};(-42,-4)*{};(-39,-1)*{}**\dir{-};(-38,-4)*\xycircle(0.5,0.5){}*\frm{*};(-38,-4)*{};(-41,-1)*{}**\dir{-};(-38,-4)*{};(-39,-1)*{}**\dir{-};(-38,-4)*{};(-37,-1)*{}**\dir{-};(-38,-4)*{};(-35,-1)*{}**\dir{-};(-34,-4)*\xycircle(0.5,0.5){}*\frm{*};(-34,-4)*{};(-37,-1)*{}**\dir{-};(-34,-4)*{};(-35,-1)*{}**\dir{-};(-34,-4)*{};(-33,-1)*{}**\dir{-};(-34,-4)*{};(-31,-1)*{}**\dir{-};(-30,-4)*\xycircle(0.5,0.5){}*\frm{*};(-30,-4)*{};(-33,-1)*{}**\dir{-};(-30,-4)*{};(-31,-1)*{}**\dir{-};(-30,-4)*{};(-29,-1)*{}**\dir{-};(-30,-4)*{};(-27,-1)*{}**\dir{-};(-26,-4)*\xycircle(0.5,0.5){}*\frm{*};(-26,-4)*{};(-29,-1)*{}**\dir{-};(-26,-4)*{};(-27,-1)*{}**\dir{-};(-26,-4)*{};(-25,-1)*{}**\dir{-};(-26,-4)*{};(-23,-1)*{}**\dir{-};(-22,-4)*\xycircle(0.5,0.5){}*\frm{*};(-22,-4)*{};(-25,-1)*{}**\dir{-};(-22,-4)*{};(-23,-1)*{}**\dir{-};(-22,-4)*{};(-21,-1)*{}**\dir{-};(-22,-4)*{};(-19,-1)*{}**\dir{-};(-18,-4)*\xycircle(0.5,0.5){}*\frm{*};(-18,-4)*{};(-21,-1)*{}**\dir{-};(-18,-4)*{};(-19,-1)*{}**\dir{-};(-18,-4)*{};(-17,-1)*{}**\dir{-};(-18,-4)*{};(-15,-1)*{}**\dir{-};(-14,-4)*\xycircle(0.5,0.5){}*\frm{*};(-14,-4)*{};(-17,-1)*{}**\dir{-};(-14,-4)*{};(-15,-1)*{}**\dir{-};(-14,-4)*{};(-13,-1)*{}**\dir{-};(-14,-4)*{};(-11,-1)*{}**\dir{-};(-10,-4)*\xycircle(0.5,0.5){}*\frm{*};(-10,-4)*{};(-13,-1)*{}**\dir{-};(-10,-4)*{};(-11,-1)*{}**\dir{-};(-10,-4)*{};(-9,-1)*{}**\dir{-};(-10,-4)*{};(-7,-1)*{}**\dir{-};(-6,-4)*\xycircle(0.5,0.5){}*\frm{*};(-6,-4)*{};(-9,-1)*{}**\dir{-};(-6,-4)*{};(-7,-1)*{}**\dir{-};(-6,-4)*{};(-5,-1)*{}**\dir{-};(-6,-4)*{};(-3,-1)*{}**\dir{-};(-2,-4)*\xycircle(0.5,0.5){}*\frm{*};(-2,-4)*{};(-5,-1)*{}**\dir{-};(-2,-4)*{};(-3,-1)*{}**\dir{-};(-2,-4)*{};(-1,-1)*{}**\dir{-};(-2,-4)*{};(1,-1)*{}**\dir{-};(2,-4)*\xycircle(0.5,0.5){}*\frm{*};(2,-4)*{};(-1,-1)*{}**\dir{-};(2,-4)*{};(1,-1)*{}**\dir{-};(2,-4)*{};(3,-1)*{}**\dir{-};(2,-4)*{};(5,-1)*{}**\dir{-};(6,-4)*\xycircle(0.5,0.5){}*\frm{*};(6,-4)*{};(3,-1)*{}**\dir{-};(6,-4)*{};(5,-1)*{}**\dir{-};(6,-4)*{};(7,-1)*{}**\dir{-};(6,-4)*{};(9,-1)*{}**\dir{-};(10,-4)*\xycircle(0.5,0.5){}*\frm{*};(10,-4)*{};(7,-1)*{}**\dir{-};(10,-4)*{};(9,-1)*{}**\dir{-};(10,-4)*{};(11,-1)*{}**\dir{-};(10,-4)*{};(13,-1)*{}**\dir{-};(14,-4)*\xycircle(0.5,0.5){}*\frm{*};(14,-4)*{};(11,-1)*{}**\dir{-};(14,-4)*{};(13,-1)*{}**\dir{-};(14,-4)*{};(15,-1)*{}**\dir{-};(14,-4)*{};(17,-1)*{}**\dir{-};(18,-4)*\xycircle(0.5,0.5){}*\frm{*};(18,-4)*{};(15,-1)*{}**\dir{-};(18,-4)*{};(17,-1)*{}**\dir{-};(18,-4)*{};(19,-1)*{}**\dir{-};(18,-4)*{};(21,-1)*{}**\dir{-};(22,-4)*\xycircle(0.5,0.5){}*\frm{*};(22,-4)*{};(19,-1)*{}**\dir{-};(22,-4)*{};(21,-1)*{}**\dir{-};(22,-4)*{};(23,-1)*{}**\dir{-};(22,-4)*{};(25,-1)*{}**\dir{-};(26,-4)*\xycircle(0.5,0.5){}*\frm{*};(26,-4)*{};(23,-1)*{}**\dir{-};(26,-4)*{};(25,-1)*{}**\dir{-};(26,-4)*{};(27,-1)*{}**\dir{-};(26,-4)*{};(29,-1)*{}**\dir{-};(30,-4)*\xycircle(0.5,0.5){}*\frm{*};(30,-4)*{};(27,-1)*{}**\dir{-};(30,-4)*{};(29,-1)*{}**\dir{-};(30,-4)*{};(31,-1)*{}**\dir{-};(30,-4)*{};(33,-1)*{}**\dir{-};(34,-4)*\xycircle(0.5,0.5){}*\frm{*};(34,-4)*{};(31,-1)*{}**\dir{-};(34,-4)*{};(33,-1)*{}**\dir{-};(34,-4)*{};(35,-1)*{}**\dir{-};(34,-4)*{};(37,-1)*{}**\dir{-};(38,-4)*\xycircle(0.5,0.5){}*\frm{*};(38,-4)*{};(35,-1)*{}**\dir{-};(38,-4)*{};(37,-1)*{}**\dir{-};(38,-4)*{};(39,-1)*{}**\dir{-};(38,-4)*{};(41,-1)*{}**\dir{-};(42,-4)*\xycircle(0.5,0.5){}*\frm{*};(42,-4)*{};(39,-1)*{}**\dir{-};(42,-4)*{};(41,-1)*{}**\dir{-};(42,-4)*{};(43,-1)*{}**\dir{-};(42,-4)*{};(45,-1)*{}**\dir{-};(46,-4)*\xycircle(0.5,0.5){}*\frm{*};(46,-4)*{};(43,-1)*{}**\dir{-};(46,-4)*{};(45,-1)*{}**\dir{-};(46,-4)*{};(47,-1)*{}**\dir{-};(46,-4)*{};(49,-1)*{}**\dir{-};(49,-3)*{};(47,-1)*{}**\dir{-};(49,-1)*{};(49,-1)*{}**\dir{-};(-49,-7)*{};(-46,-4)*{}**\dir{-};(-44,-10)*\xycircle(0.5,0.5){}*\frm{*};(-44,-10)*{};(-49,-5)*{}**\dir{-};(-44,-10)*{};(-46,-4)*{}**\dir{-};(-44,-10)*{};(-42,-4)*{}**\dir{-};(-44,-10)*{};(-38,-4)*{}**\dir{-};(-36,-10)*\xycircle(0.5,0.5){}*\frm{*};(-36,-10)*{};(-42,-4)*{}**\dir{-};(-36,-10)*{};(-38,-4)*{}**\dir{-};(-36,-10)*{};(-34,-4)*{}**\dir{-};(-36,-10)*{};(-30,-4)*{}**\dir{-};(-28,-10)*\xycircle(0.5,0.5){}*\frm{*};(-28,-10)*{};(-34,-4)*{}**\dir{-};(-28,-10)*{};(-30,-4)*{}**\dir{-};(-28,-10)*{};(-26,-4)*{}**\dir{-};(-28,-10)*{};(-22,-4)*{}**\dir{-};(-20,-10)*\xycircle(0.5,0.5){}*\frm{*};(-20,-10)*{};(-26,-4)*{}**\dir{-};(-20,-10)*{};(-22,-4)*{}**\dir{-};(-20,-10)*{};(-18,-4)*{}**\dir{-};(-20,-10)*{};(-14,-4)*{}**\dir{-};(-12,-10)*\xycircle(0.5,0.5){}*\frm{*};(-12,-10)*{};(-18,-4)*{}**\dir{-};(-12,-10)*{};(-14,-4)*{}**\dir{-};(-12,-10)*{};(-10,-4)*{}**\dir{-};(-12,-10)*{};(-6,-4)*{}**\dir{-};(-4,-10)*\xycircle(0.5,0.5){}*\frm{*};(-4,-10)*{};(-10,-4)*{}**\dir{-};(-4,-10)*{};(-6,-4)*{}**\dir{-};(-4,-10)*{};(-2,-4)*{}**\dir{-};(-4,-10)*{};(2,-4)*{}**\dir{-};(4,-10)*\xycircle(0.5,0.5){}*\frm{*};(4,-10)*{};(-2,-4)*{}**\dir{-};(4,-10)*{};(2,-4)*{}**\dir{-};(4,-10)*{};(6,-4)*{}**\dir{-};(4,-10)*{};(10,-4)*{}**\dir{-};(12,-10)*\xycircle(0.5,0.5){}*\frm{*};(12,-10)*{};(6,-4)*{}**\dir{-};(12,-10)*{};(10,-4)*{}**\dir{-};(12,-10)*{};(14,-4)*{}**\dir{-};(12,-10)*{};(18,-4)*{}**\dir{-};(20,-10)*\xycircle(0.5,0.5){}*\frm{*};(20,-10)*{};(14,-4)*{}**\dir{-};(20,-10)*{};(18,-4)*{}**\dir{-};(20,-10)*{};(22,-4)*{}**\dir{-};(20,-10)*{};(26,-4)*{}**\dir{-};(28,-10)*\xycircle(0.5,0.5){}*\frm{*};(28,-10)*{};(22,-4)*{}**\dir{-};(28,-10)*{};(26,-4)*{}**\dir{-};(28,-10)*{};(30,-4)*{}**\dir{-};(28,-10)*{};(34,-4)*{}**\dir{-};(36,-10)*\xycircle(0.5,0.5){}*\frm{*};(36,-10)*{};(30,-4)*{}**\dir{-};(36,-10)*{};(34,-4)*{}**\dir{-};(36,-10)*{};(38,-4)*{}**\dir{-};(36,-10)*{};(42,-4)*{}**\dir{-};(44,-10)*\xycircle(0.5,0.5){}*\frm{*};(44,-10)*{};(38,-4)*{}**\dir{-};(44,-10)*{};(42,-4)*{}**\dir{-};(44,-10)*{};(46,-4)*{}**\dir{-};(44,-10)*{};(49,-5)*{}**\dir{-};(49,-7)*{};(46,-4)*{}**\dir{-};(-49,-15)*{};(-44,-10)*{}**\dir{-};(-40,-22)*\xycircle(0.5,0.5){}*\frm{*};(-40,-22)*{};(-49,-13)*{}**\dir{-};(-40,-22)*{};(-44,-10)*{}**\dir{-};(-40,-22)*{};(-36,-10)*{}**\dir{-};(-40,-22)*{};(-28,-10)*{}**\dir{-};(-24,-22)*\xycircle(0.5,0.5){}*\frm{*};(-24,-22)*{};(-36,-10)*{}**\dir{-};(-24,-22)*{};(-28,-10)*{}**\dir{-};(-24,-22)*{};(-20,-10)*{}**\dir{-};(-24,-22)*{};(-12,-10)*{}**\dir{-};(-8,-22)*\xycircle(0.5,0.5){}*\frm{*};(-8,-22)*{};(-20,-10)*{}**\dir{-};(-8,-22)*{};(-12,-10)*{}**\dir{-};(-8,-22)*{};(-4,-10)*{}**\dir{-};(-8,-22)*{};(4,-10)*{}**\dir{-};(8,-22)*\xycircle(0.5,0.5){}*\frm{*};(8,-22)*{};(-4,-10)*{}**\dir{-};(8,-22)*{};(4,-10)*{}**\dir{-};(8,-22)*{};(12,-10)*{}**\dir{-};(8,-22)*{};(20,-10)*{}**\dir{-};(24,-22)*\xycircle(0.5,0.5){}*\frm{*};(24,-22)*{};(12,-10)*{}**\dir{-};(24,-22)*{};(20,-10)*{}**\dir{-};(24,-22)*{};(28,-10)*{}**\dir{-};(24,-22)*{};(36,-10)*{}**\dir{-};(40,-22)*\xycircle(0.5,0.5){}*\frm{*};(40,-22)*{};(28,-10)*{}**\dir{-};(40,-22)*{};(36,-10)*{}**\dir{-};(40,-22)*{};(44,-10)*{}**\dir{-};(40,-22)*{};(49,-13)*{}**\dir{-};(49,-15)*{};(44,-10)*{}**\dir{-};(-49,-31)*{};(-40,-22)*{}**\dir{-};(-32,-46)*\xycircle(0.5,0.5){}*\frm{*};(-32,-46)*{};(-49,-29)*{}**\dir{-};(-32,-46)*{};(-40,-22)*{}**\dir{-};(-32,-46)*{};(-24,-22)*{}**\dir{-};(-32,-46)*{};(-8,-22)*{}**\dir{-};(0,-46)*\xycircle(0.5,0.5){}*\frm{*};(0,-46)*{};(-24,-22)*{}**\dir{-};(0,-46)*{};(-8,-22)*{}**\dir{-};(0,-46)*{};(8,-22)*{}**\dir{-};(0,-46)*{};(24,-22)*{}**\dir{-};(32,-46)*\xycircle(0.5,0.5){}*\frm{*};(32,-46)*{};(8,-22)*{}**\dir{-};(32,-46)*{};(24,-22)*{}**\dir{-};(32,-46)*{};(40,-22)*{}**\dir{-};(32,-46)*{};(49,-29)*{}**\dir{-};(49,-31)*{};(40,-22)*{}**\dir{-};(-49,-63)*{};(-32,-46)*{}**\dir{-};(-28.427145923572,-81.572854076428)*{};(-49,-61)*{}**\dir{-};(-20.142381974524,-81.572854076428)*{};(-32,-46)*{}**\dir{-};(-11.857618025476,-81.572854076428)*{};(0,-46)*{}**\dir{-};(-3.5728540764279,-81.572854076428)*{};(32,-46)*{}**\dir{-};(35.572854076428,-81.572854076428)*{};(0,-46)*{}**\dir{-};(43.857618025476,-81.572854076428)*{};(32,-46)*{}**\dir{-};\endxy};\endxy\]
\caption{Partial ordered set corresponding to the binary MERA in one dimension, embedded in $\mathbb R^2$ such that the speed of light is equal to $1$ everywhere. The output lattice is the top row of dots and time flows upward. The black circles are events and the lines segments are causal links.}
\label{4to1}
\end{figure}

Therefore, one can recover the MERA simply from the causal order between the set of events, together with the dimensions of their assigned Hilbert spaces. For instance, the usual binary MERA~\cite{vidal08} is implied by the partial order shown in Figure~\ref{4to1}. Such ``causal sets'' have been studied before as discrete models of spacetime~\cite{bombelli87,hawkins03}. The idea follows from the fact that the geometry of a manifold with Lorentzian signature can be recovered exactly from the partial order between events induced by the metric, together with the volume form. The causal order directly makes sense for a discrete spacetime. For the volume form, a natural postulate is that it corresponds to the counting of events. In our case, assuming for simplicity that all events are associated with the same Hilbert space dimension, the number of events in a given spacetime region is proportional to the number of variational parameters, thank to the local representability result.  
 
In order to see what metric a MERA on a $d$-dimensional lattice may correspond to, the easiest is to first parameterize its events by coordinates in which the speed of light is constant (and equal to $1$), i.e., in a spacetime with metric 
\(
ds^2 = f(t,x_1,\dots,x_d) \bigl({-dt^2 + \sum_i dx_i^2}\bigr).
\)
We suppose that each coarse-graining increases the lattice spacing by a factor $a$, and that sites at the $(k+1)$th coarse-graining step have a causal influence on the sites of the $k$th step within a radius $r a^{k}$. Then a constant speed of light (equal to $1$) is achieved by embedding the $k$th coarse-graining at time $t = -r a^k/(a-1) $. We choose it negative so that it increases chronologically with the quantum computation, outputting the final state at time $t_0=-r/(a-1)$. 
In order to determine the conformal factor $f(t,x_1,\dots,x_d)$, we postulate that in coordinates where our lattices are equally spaced in time, and renormalized, the volume form should be constant. This makes precise the idea that the number of events in a given region of spacetime should be proportional to the volume of that region. Such coordinates must be of the form $\tau = -\alpha \log[t/t_0] $ and $\zeta_i = -\beta x_i/ t$.
The constraint is then satisfied by picking $f(t) = (\alpha/t)^2$.
Also, choosing $t_0 = -r/(a-1)$ puts the output boundary $k=0$ at $\tau = 0$, and $\beta = \alpha$ normalizes the volume element. 
In the coordinates $(\tau,\zeta_1,\dots,\zeta_d)$ the metric is then
\[
ds^2 = \left({\frac{\rho^2}{\alpha^2}-1}\right) d\tau^2 - 2 \frac \rho \alpha \,d\rho \,d\tau + \sum_i d\zeta_i^2.
\]  
where $\rho^2 = \sum_i \zeta_i^2$, and the volume form has component
\(
\sqrt{|\det g|} = 1.
\)
In the conformally flat coordinates this is
\[
ds^2 = \left({\frac{\alpha}{t}}\right)^2 \bigl({-dt^2 + \sum_i dx_i^2}\bigr).
\]
This metric is that of de Sitter space.
Another common coordinate system is given by the time coordinate $\tau$ together with $\xi_i = -\alpha x_i / t_0$, so that
\[
ds^2 = -d\tau^2 + e^{2 \tau / \alpha}\sum_i d\xi_i^2.
\]

\begin{figure} 
\begin{center}
\includegraphics[
scale=1.2]{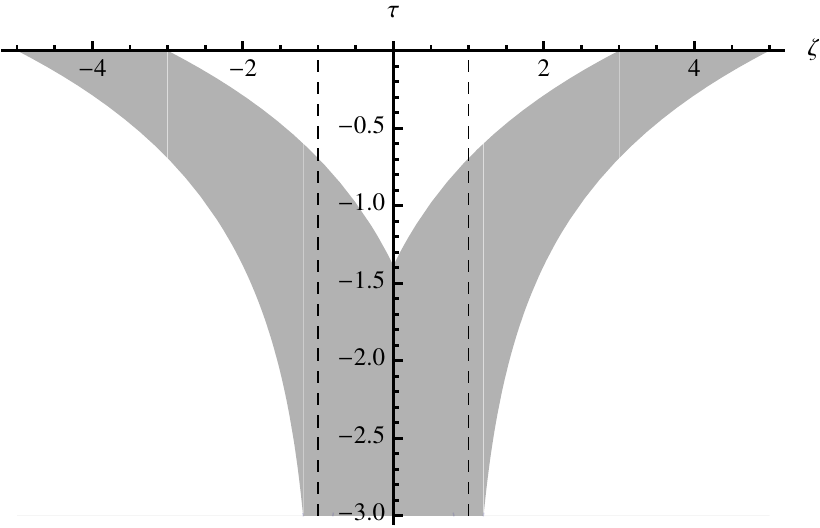}
\end{center}
\caption{The shaded area is the causal past of two disconnected regions of the $\tau=0$ spacelike surface in the static coordinates $(\tau,\zeta)$. The dashed lines indicate the horizon at $|\zeta| = \alpha = 1$.}
\label{correlations}
\end{figure}
 
Basic properties of the MERA can be deduced from considering past lightcones in the coordinates $(\tau,\zeta_1,\dots,\zeta_d)$ with constant volume form. 
The lightlike worldlines can be deduced by applying the coordinate change to the Minkowski ones. They are all of the form
\[
\zeta_i(\tau)= - \alpha u_i(1-e^{\tau/\alpha}) + \zeta_i(0)e^{\tau/\alpha}
\]
where $u_i$ a unit vector. 
We see that the causal past of any bounded region of size $L$ converges in the past $\tau/\alpha \rightarrow -\infty$ to the ball of radius $\alpha$ (the cosmological horizon), which contains a fixed number of lattice sites at any given time. This is precisely the feature of a MERA which allows for the efficient computation of local expectation values, since computing a reduced density matrix means simulating the quantum dynamics within the horizon which contains a bounded number of sites independent of the lattice size. 
Figure \ref{correlations} illustrates this phenomenon. It shows the causal past associated with the computation of the correlation function between two local observables for $d=1$. The log-dimension of the tensors that one needs to contract at each time step is proportional to the spacelike volume of the shaded area at that time. 

\prlsection{Continuous MERA}

This analysis yields a possible approach to building a continuous MERA, to serve as an ansatz for the state of quantum field theories. The field state to be represented lives on the spacelike boundary at $\tau = 0$, and is the state produced by the evolution of {\em some} matter field inside de Sitter space.
The boundary state is a function of the actual matter Hamiltonian, which plays the same role as the gate parameters in the discrete case. Indeed, the fact that the field ``lives'' in de Sitter space simply means that its dynamics respects the corresponding causality conditions, which is usually formalized by saying that the field operators evaluated at spacelike-separated events must commute (or anticommute if they are fermions). 
Just as in the discrete case, one must also choose an initial state inside the horizon at a sufficiently early time. For instance, starting at $\tau = -\alpha \log((L + 2 \alpha)/4\alpha)$ allows for correlations length up to $L$. However, this initial value problem is not the only way of using the ansatz. For instance, one may also choose a stationary matter Hamiltonian, i.e., symmetric under translation in $\tau$ with $\zeta$ constant---which mirrors the strategy of choosing the gates to be the same at each step in the discrete case---and then define the boundary state to be a stationary state. This is expected to yield long range correlations.

Of course, the discrete MERA can also be used to attempt to approximate the state of quantum fields provided the QFT is first discretized. However, working directly in the continuum can present many advantages, such as the ability to deal with more symmetries. A naive re-discretization would yield back a MERA, but one could conceive other ways of producing a numerical algorithm such as, for instance, by applying a covariant cutoff which preserves the continuous symmetries such as in Ref.~\cite{kempf01}.

A continuous MERA, or cMERA, was proposed recently by Haegeman {\it et al.}~\cite{haegeman11}. We will see that our proposal constitutes in some sense a subset of theirs. 
In order to compare them, we consider the simplest example of a quantum scalar field: with Lagrangian density
\[
\mathcal L = \frac 1 2 \sqrt{|\det g|} [ - g^{\mu\nu}\phi_{,\mu}\phi_{,\nu} - V(\phi) ].
\]

To be clear, it is to be understood that this is {\em not} the boundary theory that the cMERA is meant to help us solve, but instead a particular choice of the ansatz's variational parameters (with some freedom left in the choice of $V$, which could depend on both space and time). Hence we may say that this Lagrangian defines the {\em bulk} theory. The ansatz itself is reflected in the fact that this field, once properly quantized, respects de Sitter causality. 

The proposal in Ref.~\cite{haegeman11} does not explicitly require causality constraints. Instead, emphasis is put on imposing an ultraviolet cutoff. Given a local Hamiltonian, this would automatically create approximate causality constraints due to the Lieb-Robinson bounds which limit the speed at which information propagates for any dynamics generated by a local Hamiltonian on a lattice. 
This is not needed in our proposal which comes equipped with exact causality constraints. Instead, the spatial components of the metric yield a natural concept of length at each ``scale'', i.e. each time $\tau$. That is, if a cutoff were to be applied, it would be measured relative to our metric. For instance, the binary MERA corresponds to a cutoff $\Delta \zeta \simeq 2\alpha/3$. 
This tells us that the coordinate system used in Ref.~\cite{haegeman11} must be compared to our static coordinate system $(\tau,\zeta_1,\dots)$, given that they use the same cutoff at each time step. The Lagrangian $\mathcal L$ can be quantized by standard methods. If $\hat \pi(\zeta_1,\dots,\zeta_d)$ are the canonical conjugates to the field operators $\hat \phi(\zeta_1,\dots,\zeta_d)$, then we obtain the Hamiltonian $H = K + L$, where
\[
K = \frac 1 2 \int d^d\zeta \Bigl[{ \hat \pi^2 + \sum_i (\hat \phi_{,i})^2  + V(\hat \phi)}\Bigr]
\]
is the Hamiltonian for the field on flat spacetime and
\[
L = -\frac 1 \alpha \int d^d\zeta \sum_i \frac 1 2 \Bigl[{ \hat \pi \zeta_i \hat \phi_{,i} + \zeta_i \hat \phi_{,i} \hat \pi }\Bigr].
\]
generates the expansion of space. One can easily check that, with canonical commutation relations between $\hat \pi$ and $\hat \phi$, this yields the right Heisenberg equations of motion. This structure for $H$ is compatible with that proposed by Haegeman {\it et al}. 

\section{Continuum limit of MERA}

At this point, it is worth making some observations concerning the continuum limit of a discrete MERA. One possibility (which corresponds to what is done in Ref.~\cite{pfeifer09}) would imply keeping the same infinite network all along, but virtually rescaling the length and time coordinates $x$ and $t$ as if we were looking at the network from far away. The horizon length is sent to zero in this scheme which, therefore, cannot lead to a proper de Sitter limit. This makes sense if this is to lead to a scale-invariant boundary state in the limit. Indeed, the presence of a finite horizon introduces a characteristic lengthscale which would be of no use for describing a scale invariant state. 

Note that the horizon size should not be thought of as a correlation length, which would be better encoded in a time-dependent (i.e. scale-dependent) bulk Hamiltonian. Instead, the state inside the horizon would look like it has been produced by some dynamics on a flat spacetime, and may exhibit strong entanglement that scales as the volume of a region.

In order to obtain de Sitter space from a discrete MERA, one may imagine a network which is not defined exactly by de Sitter causality but instead by a causal structure resembling our universe: with an initial period of exponential expansion (i.e., de Sitter geometry) which transitions to a smaller (polynomial) rate of expansion until it outputs the present state of the universe. If the polynomial period is long enough, the de Sitter horizon can be macroscopically large compared to the lattice spacing, so that the intermediate state at the end of the inflationary period appears being produced by a continuous de Sitter space. Such a network would look like that of Fig. 1e in Ref.~\cite{dawson08}. 

The final period undergoing a sub-exponential expansion can be thought of as an example of a MERA where, as we move back in ``time'' from the output boundary, the dimension of the Hilbert spaces associated with the lattice sites increases for a while, which is actually what is done in simulations~\cite{evenbly09b}. The difference however is that, instead of simply increasing the bond dimension, this involves actually adding extra lattice sites within the light cone. This means that these larger isometries are forced to satisfy finer causality constraints which break them into smaller local pieces, hence decreasing the number of variational parameters compared to a simple increase in dimension (although this does not reduce the computational time involved in contracting these tensors).

\prlsection{AdS-CFT correspondence}

Previous works have compared MERA to a different continuum theory, namely anti de Sitter (AdS) space. 
The idea being that MERA may be a concrete embodiment of the AdS-CFT correspondence for its ability to represent critical scale-invariant states and the similarity in which entanglement entropy is calculated using a minimal surface in the bulk~\cite{swingle09, evenbly11}. This suggests an interpretation where the MERA circuit represents a field theory on AdS spacetime, whereas the state it describes is a CFT on a {\em timelike} boundary. 
The AdS metric is
\[
ds^2 = \Bigl({ \frac{\alpha}{x_1} }\Bigr)^2( -dt^2 + \sum_i dx_i^2 )
\]
and the boundary considered corresponds to a fixed value of the coordinate $x_1$.
If we make the signature Euclidean by changing the sign of $dt^2$---a common trick used in QFT---then both the AdS metric and the de Sitter metric become the same hyperbolic metric. Furthermore, the boundary that we have been considering matches the one considered in the AdS-CFT correspondence. 

This shows that the Euclidean form of our field theory lives precisely on the same spacetime, and with the same boundary, as the Euclidean form of AdS spacetime. It is interesting that the argument we have used to relate MERA to hyperbolic geometry appears to be distinct from previous arguments in the context of AdS-CFT~\cite{swingle09}, which are based essentially on an analogy between the way entanglement entropy is calculated in AdS-CFT and MERA.

Going back to a Lorentzian metric signature, we see that the AdS boundary being timelike, it naturally contains a time direction in which the boundary theory is to evolve. In this sense the AdS time is the physical time, whereas the role of scale is played by the space coordinate $x_1$. 
In our de Sitter picture, however, the Lorentzian time $t$ is the scale parameter and the boundary is spacelike. This does not mean that the ansatz cannot be used to describe a time evolving state. Typically, conformal field theories are described in the Euclidean form, and hence would naturally live on our spacelike boundary. This is in fact the standard approach used with the discrete MERA~\cite{pfeifer09}. Additionally, it is possible in principle to simulate the time evolution of a state within the MERA variational class of states, as proposed in Ref.~\cite{rizzi08}. In this case, physical time is a new parameter in terms of which the ansatz's parameters are varying.

\prlsection{Causality and locality}
We now sketch the proof of the statement represented by Equ.~\ref{step}. Namely, that the set of isometries which are purely causal with respect to the causality relation represented on the left-hand side is equal to the set of isometries which can be formed by replacing each box of the right hand side by an isometry (without constraint on the dimension of the Hilbert spaces associated with the middle wires).
First we consider a unitary map $U$ whose inputs are grouped into systems $A$ and $B$, and outputs are grouped into systems $A'$ and $B'$, with the constraint that $B$ cannot influence $A'$. This means that in the Heisenberg picture, any operator $X$ acting on system $A'$ is mapped to an operator $Y$ acting only on system $A$, i.e. $U^\dagger (X_{A'} \otimes \one_{B'}) U = Y_{A} \otimes \one_{B}$, which can be rewritten as
\begin{equation}
\label{found}
(X_{A'} \otimes \one_{B'}) \,U = U\, (Y_{A} \otimes \one_{B}).
\end{equation}
This implies that for any pure state $\ket x$ of $B$, 
\[
Y = (\one \otimes \bra x) \,U^\dagger\, (X \otimes \one)\, U \,(\one \otimes \ket x).
\]
 The trick, inspired by Ref.~\cite{arrighi10}, is to replace the local operator $X$ by a swap between $A'$ and a new system $C$ in Equation~\eqref{found}. If we initialize the system $C$ to an arbitrary state $\ket y$ and trace it out after the action of $U$ and the swap, the left hand side becomes simply $U$. This yields the expansion
\begin{equation}
\label{tool}
\xy(0,0)*+{\xy(0,0)*{};(4,0)*{}**\dir{-};(2,-2)*{\scriptstyle A};(0,4)*{};(4,4)*{}**\dir{-};(2,6)*{\scriptstyle B};(4,-2)*{};(12,-2)*{}**\dir{-};(12,-2)*{};(12,6)*{}**\dir{-};(4,6)*{};(12,6)*{}**\dir{-};(4,-2)*{};(4,6)*{}**\dir{-};(8,2)*{U};(12,0)*{};(16,0)*{}**\dir{-};(14,-2)*{\scriptstyle A'};(12,4)*{};(16,4)*{}**\dir{-};(14,6)*{\scriptstyle B'};\endxy};\endxy=
\xy(0,0)*+{\xy(4.8,3)*{\scriptstyle x};(6,1.5)*{};(6,4.5)*{}**\crv{(3,1.5)&(3,4.5)};(6,1.5)*{};(6,4.5)*{}**\dir{-};(5.5,-3)*{};(9,-3)*{}**\dir{-};{\ar(0,-3)*{};(5.5,-3)*{}};(4.5,-5)*{\scriptstyle A};(6,3)*{};(9,3)*{}**\dir{-};(9,-4.5)*{};(18,-4.5)*{}**\dir{-};(18,-4.5)*{};(18,4.5)*{}**\dir{-};(9,4.5)*{};(18,4.5)*{}**\dir{-};(9,-4.5)*{};(9,4.5)*{}**\dir{-};(13.5,0)*{U};(21.3,0)*{\scriptstyle y};(22.5,-1.5)*{};(22.5,1.5)*{}**\crv{(19.5,-1.5)&(19.5,1.5)};(22.5,-1.5)*{};(22.5,1.5)*{}**\dir{-};(22.5,0)*{};(25.5,0)*{}**\dir{-};(22.75,3)*{};(25.5,3)*{}**\dir{-};{\ar(18,3)*{};(22.75,3)*{}};(25.5,-1.5)*{};(31.5,-1.5)*{}**\dir{-};(31.5,-1.5)*{};(31.5,4.5)*{}**\dir{-};(25.5,4.5)*{};(31.5,4.5)*{}**\dir{-};(25.5,-1.5)*{};(25.5,4.5)*{}**\dir{-};(28.5,1.5)*{U^\dagger};(31.5,3)*{};(34.5,3)*{}**\dir{-};(35.7,3)*{\scriptstyle x};(34.5,1.5)*{};(34.5,4.5)*{}**\crv{(37.5,1.5)&(37.5,4.5)};(34.5,1.5)*{};(34.5,4.5)*{}**\dir{-};(36.25,0)*{};(39,0)*{}**\dir{-};{\ar(31.5,0)*{};(36.25,0)*{}};(20.5,6)*{};(39,6)*{}**\dir{-};{\ar(0,6)*{};(20.5,6)*{}};(19.5,8)*{\scriptstyle B};(39,-1.5)*{};(48,-1.5)*{}**\dir{-};(48,-1.5)*{};(48,7.5)*{}**\dir{-};(39,7.5)*{};(48,7.5)*{}**\dir{-};(39,-1.5)*{};(39,7.5)*{}**\dir{-};(43.5,3)*{U};(48,0)*{};(51,0)*{}**\dir{-};(51,-1.5)*{};(51,1.5)*{}**\dir{-};(37.75,-3)*{};(55.5,-3)*{}**\dir{-};{\ar(18,-3)*{};(37.75,-3)*{}};(36.75,-5)*{\scriptstyle A'};(52.75,6)*{};(55.5,6)*{}**\dir{-};{\ar(48,6)*{};(52.75,6)*{}};(51.75,8)*{\scriptstyle B'};\endxy};\endxy\end{equation}
where the states $\ket x$ and $\ket y$ can be chosen arbitrarily.
This is the only algebraic property that we will need. The vertical bar ending the fourth wire means that this system is traced out: hence both sides of this equation represent channels rather than just operators. The channel on the left-hand side is just $\rho \mapsto U \rho U^\dagger$: being unitary, it is a minimal Stinespring dilation of the channel on the right hand side. From the uniqueness of the Stinespring dilation of a channel, the right-hand side has also only one Kraus operator. To find its precise form, first note that the operator
\[
\xy(0,0)*+{\xy(0,0)*{};(4,0)*{}**\dir{-};(2,2)*{\scriptstyle A};(4,-2)*{};(12,-2)*{}**\dir{-};(12,-2)*{};(12,6)*{}**\dir{-};(4,6)*{};(12,6)*{}**\dir{-};(4,-2)*{};(4,6)*{}**\dir{-};(8,2)*{V};(12,0)*{};(16,0)*{}**\dir{-};(14,-2)*{\scriptstyle A};(12,4)*{};(16,4)*{}**\dir{-};(14,6)*{\scriptstyle A'};\endxy};\endxy:=
\xy(0,0)*+{\xy(5.6,3.5)*{\scriptstyle x};(7,1.75)*{};(7,5.25)*{}**\crv{(3.5,1.75)&(3.5,5.25)};(7,1.75)*{};(7,5.25)*{}**\dir{-};(6.25,-3.5)*{};(10.5,-3.5)*{}**\dir{-};{\ar(0,-3.5)*{};(6.25,-3.5)*{}};(5.25,-5.5)*{\scriptstyle A};(7,3.5)*{};(10.5,3.5)*{}**\dir{-};(10.5,-5.25)*{};(21,-5.25)*{}**\dir{-};(21,-5.25)*{};(21,5.25)*{}**\dir{-};(10.5,5.25)*{};(21,5.25)*{}**\dir{-};(10.5,-5.25)*{};(10.5,5.25)*{}**\dir{-};(15.75,0)*{U};(24.85,0)*{\scriptstyle y};(26.25,-1.75)*{};(26.25,1.75)*{}**\crv{(22.75,-1.75)&(22.75,1.75)};(26.25,-1.75)*{};(26.25,1.75)*{}**\dir{-};(26.25,0)*{};(29.75,0)*{}**\dir{-};(26.375,3.5)*{};(29.75,3.5)*{}**\dir{-};{\ar(21,3.5)*{};(26.375,3.5)*{}};(29.75,-1.75)*{};(36.75,-1.75)*{}**\dir{-};(36.75,-1.75)*{};(36.75,5.25)*{}**\dir{-};(29.75,5.25)*{};(36.75,5.25)*{}**\dir{-};(29.75,-1.75)*{};(29.75,5.25)*{}**\dir{-};(33.25,1.75)*{U^\dagger};(36.75,3.5)*{};(40.25,3.5)*{}**\dir{-};(41.65,3.5)*{\scriptstyle x};(40.25,1.75)*{};(40.25,5.25)*{}**\crv{(43.75,1.75)&(43.75,5.25)};(40.25,1.75)*{};(40.25,5.25)*{}**\dir{-};(36.875,-3.5)*{};(50.75,-3.5)*{}**\dir{-};{\ar(21,-3.5)*{};(36.875,-3.5)*{}};(35.875,-5.5)*{\scriptstyle A};(44.75,0)*{};(50.75,0)*{}**\dir{-};{\ar(36.75,0)*{};(44.75,0)*{}};(43.75,2)*{\scriptstyle A'};\endxy};\endxy\]
is an isometry as can be checked by tracing out $A'$ and $B'$ on both sides of Equation~\eqref{tool}. Then the Stinespring dilation theorem tells us that there is an isometry (here just a ket) $\ket \psi$ embedding $\mathbb C$ into the Hilbert space of the system $A'$ such that
\[
\xy(0,0)*+{\xy(5,0)*{};(8,0)*{}**\dir{-};{\ar(0,0)*{};(5,0)*{}};(4,-2)*{\scriptstyle A};(5,8)*{};(8,8)*{}**\dir{-};{\ar(0,8)*{};(5,8)*{}};(4,10)*{\scriptstyle B};(8,-2)*{};(20,-2)*{}**\dir{-};(20,-2)*{};(20,10)*{}**\dir{-};(8,10)*{};(20,10)*{}**\dir{-};(8,-2)*{};(8,10)*{}**\dir{-};(14,4)*{U};(24.4,4)*{\scriptstyle \psi};(26,2)*{};(26,6)*{}**\crv{(22,2)&(22,6)};(26,2)*{};(26,6)*{}**\dir{-};(26,0)*{};(30,0)*{}**\dir{-};{\ar(20,0)*{};(26,0)*{}};(25,-2)*{\scriptstyle A'};(26,4)*{};(30,4)*{}**\dir{-};(28,2)*{\scriptstyle A'};(26,8)*{};(30,8)*{}**\dir{-};{\ar(20,8)*{};(26,8)*{}};(25,10)*{\scriptstyle B'};\endxy};\endxy=
\xy(0,0)*+{\xy(5,0)*{};(8,0)*{}**\dir{-};{\ar(0,0)*{};(5,0)*{}};(8,-2)*{};(16,-2)*{}**\dir{-};(16,-2)*{};(16,6)*{}**\dir{-};(8,6)*{};(16,6)*{}**\dir{-};(8,-2)*{};(8,6)*{}**\dir{-};(12,2)*{V};(16,4)*{};(20,4)*{}**\dir{-};(11,8)*{};(20,8)*{}**\dir{-};{\ar(0,8)*{};(11,8)*{}};(10,10)*{\scriptstyle B};(20,2)*{};(28,2)*{}**\dir{-};(28,2)*{};(28,10)*{}**\dir{-};(20,10)*{};(28,10)*{}**\dir{-};(20,2)*{};(20,10)*{}**\dir{-};(24,6)*{U};(25,0)*{};(32,0)*{}**\dir{-};{\ar(16,0)*{};(25,0)*{}};(24,-2)*{\scriptstyle A'};(28,8)*{};(32,8)*{}**\dir{-};(30,10)*{\scriptstyle B'};(28,4)*{};(32,4)*{}**\dir{-};(30,2)*{\scriptstyle A'};\endxy};\endxy\]
It follows that we can replace the channel $\chan N(\rho) := \tr_{A'}U \rho U^\dagger$ in Equ.~\eqref{tool} by $\rho \mapsto X \rho X^\dagger$, with $X := (\one_{B'} \otimes \bra \psi_{A'}) U$. Furthermore, since the whole expression must be unitary, and hence trace-preserving, the operator $X$ is isometric when restricted to its possible inputs in the circuit, and can therefore be replaced by an isometry. 

This can be used to parameterize the classes of unitary maps causal with respect to a relation like that of Equ.~\eqref{step} as follows: we start by grouping all the inputs (resp. outputs) which have the set of children (resp. parents) to obtain a new causal relation on the grouped systems.
If the resulting graph is such that removing one particular input $A$ breaks it into two independent parts, then the remaining inputs and outputs can be grouped so as to satisfy the causality relation
\[
\xy(0,0)*+{\xy
(-5,0)*\xycircle(0.5,0.5){}*\frm{*};
(0,0)*\xycircle(0.5,0.5){}*\frm{*};
(5,0)*\xycircle(0.5,0.5){}*\frm{*};
(-5,5)*\xycircle(0.5,0.5){}*\frm{*};
(5,5)*\xycircle(0.5,0.5){}*\frm{*};
(-5,0)*{};(-5,5)*{}**\dir{-};
(5,0)*{};(5,5)*{}**\dir{-};
(0,0)*{};(-5,5)*{}**\dir{-};
(0,0)*{};(5,5)*{}**\dir{-};
(0,-3)*{A};
\endxy};\endxy
\]
This represents two causality constraints (i.e. missing links). By applying the instance of Equ.~\ref{tool} allowed by one of the constraint, and then again on the first instance of $U$ in the circuit for the other constraint, we obtain that
\[
\xy(0,0)*+{\xy(-6,3)*{};(-6,4)*{}**\dir{-};{\ar(-6,0)*{};(-6,3)*{}};(0,3)*{};(0,4)*{}**\dir{-};{\ar(0,0)*{};(0,3)*{}};(6,3)*{};(6,4)*{}**\dir{-};{\ar(6,0)*{};(6,3)*{}};(-7,4)*{};(-7,18)*{}**\dir{-};(-7,18)*{};(7,18)*{}**\dir{-};(7,4)*{};(7,18)*{}**\dir{-};(-7,4)*{};(7,4)*{}**\dir{-};(0,11)*{U};(-6,21)*{};(-6,22)*{}**\dir{-};{\ar(-6,18)*{};(-6,21)*{}};(6,21)*{};(6,22)*{}**\dir{-};{\ar(6,18)*{};(6,21)*{}};\endxy};\endxy 
= \xy(0,0)*+{\xy(0,0)*{};(0,2)*{}**\dir{-};(-5,2)*{};(-5,12)*{}**\dir{-};(-5,12)*{};(5,12)*{}**\dir{-};(5,2)*{};(5,12)*{}**\dir{-};(-5,2)*{};(5,2)*{}**\dir{-};(0,7)*{V};(-6,8)*{};(-6,14)*{}**\dir{-};{\ar(-6,0)*{};(-6,8)*{}};(-4,12)*{};(-4,14)*{}**\dir{-};(-2,12)*{};(-2,14)*{}**\dir{-};(-7,14)*{};(-7,20)*{}**\dir{-};(-7,20)*{};(-1,20)*{}**\dir{-};(-1,14)*{};(-1,20)*{}**\dir{-};(-7,14)*{};(-1,14)*{}**\dir{-};(-4,17)*{U};(2,12)*{};(2,14)*{}**\dir{-};(4,12)*{};(4,14)*{}**\dir{-};(6,8)*{};(6,14)*{}**\dir{-};{\ar(6,0)*{};(6,8)*{}};(1,14)*{};(1,20)*{}**\dir{-};(1,20)*{};(7,20)*{}**\dir{-};(7,14)*{};(7,20)*{}**\dir{-};(1,14)*{};(7,14)*{}**\dir{-};(4,17)*{U};(-2,20)*{};(-2,22)*{}**\dir{-};(-3,22)*{};(-1,22)*{}**\dir{-};(2,20)*{};(2,22)*{}**\dir{-};(1,22)*{};(3,22)*{}**\dir{-};(-6,23.5)*{};(-6,25)*{}**\dir{-};{\ar(-6,20)*{};(-6,23.5)*{}};(6,23.5)*{};(6,25)*{}**\dir{-};{\ar(6,20)*{};(6,23.5)*{}};\endxy};\endxy\]
for some isometry $V$.
This scheme can be applied recursively on the remaining copies of $U$, until the circuit respects all the causality constraints. 

If we lift the restriction that our computational step be unitary, and assume instead that it is an isometry (as is required for a MERA) or more generally a quantum channel, then we demand that it can be represented by a unitary interaction with a local environment, such that the unitary map respects the same causal relation. We also require that the environment's initial state is separable. We can then apply our procedure to this unitary map to show that it has a local representation. In this way, one obtains the result express in Equ.~\eqref{step}.
This method also works for the ternary MERA, showing that
  \[
\cdots
\xy(0,0)*+{\xy
  (-15,0)*\xycircle(0.5,0.5){}*\frm{*};
  (-10,0)*\xycircle(0.5,0.5){}*\frm{*};
  (-5,0)*\xycircle(0.5,0.5){}*\frm{*};
  (0,0)*\xycircle(0.5,0.5){}*\frm{*};
  (5,0)*\xycircle(0.5,0.5){}*\frm{*};
  (10,0)*\xycircle(0.5,0.5){}*\frm{*};
  (15,0)*\xycircle(0.5,0.5){}*\frm{*};
  (-15,-7.5)*\xycircle(0.5,0.5){}*\frm{*};
  (0,-7.5)*\xycircle(0.5,0.5){}*\frm{*};
  (15,-7.5)*\xycircle(0.5,0.5){}*\frm{*};
  (-15,-7.5)*{};(-15,0)*{}**\dir{-};
  (-15,-7.5)*{};(-10,0)*{}**\dir{-};
  (-15,-7.5)*{};(-5,0)*{}**\dir{-};
  (0,-7.5)*{};(-10,0)*{}**\dir{-};
  (0,-7.5)*{};(-5,0)*{}**\dir{-};
  (0,-7.5)*{};(0,0)*{}**\dir{-};
  (0,-7.5)*{};(5,0)*{}**\dir{-};
  (0,-7.5)*{};(10,0)*{}**\dir{-};
  (15,-7.5)*{};(15,0)*{}**\dir{-};
  (15,-7.5)*{};(10,0)*{}**\dir{-};
  (15,-7.5)*{};(5,0)*{}**\dir{-};
  ;\endxy};\endxy \cdots
 \;\; = \;
  \cdots
  \xy(0,0)*+{\xy(0,0)*{};(0,2)*{}**\dir{-};(-3,2)*{};(-3,8)*{}**\dir{-};(-3,8)*{};(3,8)*{}**\dir{-};(3,2)*{};(3,8)*{}**\dir{-};(-3,2)*{};(3,2)*{}**\dir{-};(0,5)*{{}};(8,0)*{};(8,2)*{}**\dir{-};(5,2)*{};(5,8)*{}**\dir{-};(5,8)*{};(11,8)*{}**\dir{-};(11,2)*{};(11,8)*{}**\dir{-};(5,2)*{};(11,2)*{}**\dir{-};(8,5)*{{}};(16,0)*{};(16,2)*{}**\dir{-};(13,2)*{};(13,8)*{}**\dir{-};(13,8)*{};(19,8)*{}**\dir{-};(19,2)*{};(19,8)*{}**\dir{-};(13,2)*{};(19,2)*{}**\dir{-};(16,5)*{{}};(10,8)*{};(10,10)*{}**\dir{-};(14,8)*{};(14,10)*{}**\dir{-};(9,10)*{};(9,16)*{}**\dir{-};(9,16)*{};(15,16)*{}**\dir{-};(15,10)*{};(15,16)*{}**\dir{-};(9,10)*{};(15,10)*{}**\dir{-};(12,13)*{{}};(2,8)*{};(2,10)*{}**\dir{-};(6,8)*{};(6,10)*{}**\dir{-};(1,10)*{};(1,16)*{}**\dir{-};(1,16)*{};(7,16)*{}**\dir{-};(7,10)*{};(7,16)*{}**\dir{-};(1,10)*{};(7,10)*{}**\dir{-};(4,13)*{{}};(8,14)*{};(8,18)*{}**\dir{-};{\ar(8,8)*{};(8,14)*{}};(-2,14)*{};(-2,18)*{}**\dir{-};{\ar(-2,8)*{};(-2,14)*{}};(2,16)*{};(2,18)*{}**\dir{-};(6,16)*{};(6,18)*{}**\dir{-};(10,16)*{};(10,18)*{}**\dir{-};(14,16)*{};(14,18)*{}**\dir{-};(18,14)*{};(18,18)*{}**\dir{-};{\ar(18,8)*{};(18,14)*{}};\endxy};\endxy  \cdots,
  \]  
with the same disclaimer about the fact that the dimensionality of intermediate wires are not constrained, but limited by the fact that the boxes must represent isometries.  

As mentioned in the introduction, for more general causality relations, in particular as applied to higher-dimensional lattices, one must use the general prescriptions introduced in Ref.~\cite{arrighi10}.

\section*{Acknowledgment}

The author is grateful to Tobias Osborne and Guifre Vidal for discussions about this work. This work was supported by the cluster of excellence EXC 201 “Quantum Engineering and Space-Time Research”.

\section*{References}
\bibliographystyle{unsrt}

\bibliography{desitter} 

\end{document}